\documentclass[aps,pra,superscriptaddress,twocolumn]{revtex4}
\usepackage{amsmath,epsfig,mathrsfs,graphicx,amssymb,color}
\usepackage[T1]{fontenc}
\usepackage{t1enc}
\usepackage{hyperref}
\usepackage{array}
\usepackage{booktabs}

\newcommand{\g}[1]{{\boldsymbol #1}}

\def\be#1\ee{\begin{equation}#1\end{equation}}
\newcommand{\ba}{\begin{eqnarray} }
\newcommand{\ea}{\end{eqnarray} }

\def\mb{\begin{pmatrix}}
\def\me{\end{pmatrix}}
\def\be#1\ee{\begin{equation}#1\end{equation}}

\begin{document} 

\title{Optimal discrimination between real and complex quantum theories}

\author{Adam Bednorz}
\affiliation{Faculty of Physics, University of Warsaw, ul. Pasteura 5, PL02-093 Warsaw, Poland}
\email{Adam.Bednorz@fuw.edu.pl}
\author{Josep Batle}
\affiliation{CRISP-Centre de Recerca Independent de sa Pobla, C. Alb\'eniz 12, 07420 sa Pobla, Balearic Islands, Spain}
\email{batlequantum@gmail.com}

\begin{abstract}
We find the minimal number of settings to test quantum theory based on real numbers, assuming
separability of the sources, modifying the recent proposal [M.-O. Renou et al., Nature 600,  625 (2021)].
The test needs only three settings for observers $A$ and $C$, but the ratio of complex to real maximum is smaller than in the existing proposal.
We also found that two settings and two outcomes for both observes are insufficient.

\end{abstract}

\maketitle

\section{Introduction}

Ever since the dawn of modern science, the interplay between mathematics and physics has been explored in parallel to the development of the scientific method. 
Roger Bacon firstly described mathematics as 'the door and the key to the sciences' \cite{bacon}. 
Regarding the quantification of the physical knowledge, 'The great book of nature', wrote Galileo, 'is written in mathematical language.' \cite{galileo}
And finally, quite more recently, Eugene Wigner elaborated upon 'The unreasonable effectiveness of mathematics in the natural sciences.' \cite{wigner}
In quantum mechanics, we encounter a debate not found in the classical realm, namely, how fundamental is the utilization of the field of complex numbers 
$\mathbb{C}$, as opposed to real ones $\mathbb{R}$, in the description of physical phenomena. 
In the same way that Born, Heisenberg, and ordan \cite{Heis1,Heis2}, introduced matrices in the first complete formulation of quantum mechanics, 
it was the Schr\"odinger equation that introduced explicitly $i=\sqrt{-1}$ and, therefore, complex states $\psi$ \cite{Schrod}.

Avoiding epistemological discussions such as the wave function being an element of physical reality or not \cite{reality}, 
it is no surprise that, at least experimentally, one requires the real and imaginary parts of the wave function \cite{realism}. 
Additionally, local real- or complex-valued tomography could lead to different experimental results \cite{tomography}. 
Therefore, and to no surprise, quantum physics based on complex-valued quantities is a successful theory both qualitative and quantitatively.   

Several works, notably initiated by von Neumann \cite{lit1,lit3,lit4,lit5,lit6,lit7}, 
elaborate on the possible ways of employing real numbers only for describing the same phenomena by doubling the concomitant complex 
$n$-dimensional Hilbert spaces to real-valued ones, that is, $|1\rangle,...|n\rangle \rightarrow |1\rangle,...,|n\rangle,|n+1\rangle,...,|2n\rangle$. 
This is possible for any observable or density operator given as a $n \times n$ Hermitian matrix $H=A=H_R+\,iH_I $, 
with real symmetric $H_R$ and antisymmetric $H_I$ can be regarded as an equivalent to the real, symmetric problem
\be
\begin{pmatrix}
  H_R & -H_I \\
  H_I & H_R
\end{pmatrix}.
\ee
In this fashion, each state $|\psi\rangle=\sum_k (\psi_{kR}+i\psi_{kI})|k\rangle$, with real $\psi_{kR,I}$ is replaced by
 a doublet  of $\sum_k (\psi_{kR}|k\rangle+\psi_{kI}|n+k\rangle)$,  $\sum_k (\psi_{kI}|k\rangle-\psi_{kR}|n+k\rangle)$.
Therefore  we are left also with extra degeneracy of states, which all are not normally doubled, in particular the ground state. This is not a problem
for local phenomena, but separable states consisting of several parties are doubled in each party. 
One can keep a single doublet only by extra entanglement in real space.

Recently, Renou {\it et al} \cite{cvr} developed a scheme probing this possibility, i.e. testing if the states separable in complex space need to be replaced by an entangled
state in real space. It turned out that real separability imposes additional constraints on correlations, leading to an inequality, with lower bound for
real states than for complex ones. The proposal involved three observers, $A$, $B$, and $C$, where $A$ and $B$ ($C$ and $B$)
receive qubits from the one (second) source. Then $B$ makes a single measurement with $4$ outcomes, while $A$ and $C$ make dichotomic measurements for three and six settings, respectively.
The violation of the inequality has been verified experimentally. \cite{cvre1,cvre2}

Here make an amendment to this scheme, reducing the number of settings to three for observer $C$, and constructing the corresponding witness. 
We also  provide the example with four settings, where the impossibility
of reaching the complex maximum in real space is shown. The present contribution is divided as follows. We start with the description of the setup and notation. 
Then we show the example with four settings for $C$. The later case with three settings needs a numerical search, based on a modification of the
MATLAB script published with the previous proposal \cite{cvr}. Reduction to two settings and two outcomes for both observers is impossible, as we show with the partial
help of a numerical search.
Finally, we draw some conclusions, suggesting further possible routes of optimization.

\section{Setup of the test}

\begin{figure}
\includegraphics[scale=.5]{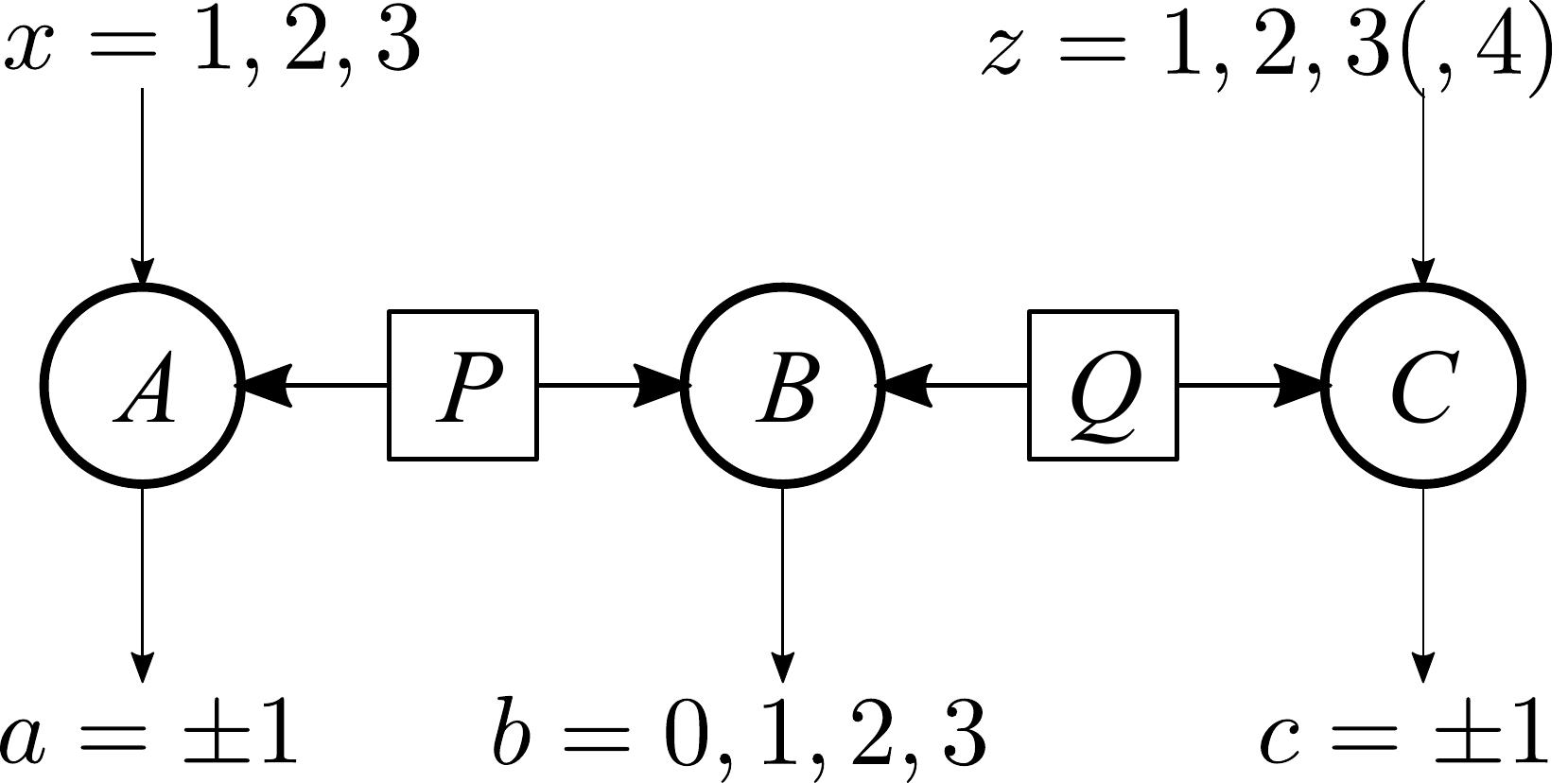}
\caption{The setup of the test. The separate sources $P$ and $Q$ generate entangled states. Central parts of these states are measured by $B$ with four possible outcomes $b$,
while the left and right parts are measured by observers $A$ and $C$, with dichotomic outcomes $a$ and $c$ for settings $x$ and $z$, respectively.
Note that the number of settings $z$ for $C$ is either $3$ or $4$.}\label{abc}
\end{figure}

The analyzed system, as in the previous work \cite{cvr}, consists of three observers $A$, $B$, and $C$, depicted in Fig. \ref{abc}. The sources $P$ and $Q$ are separable,
which is an important assumption. The observers $A$ and $C$ can choose one of three settings $x=1,2,3$ and $z=1,2,3$ (or also $4$), respectively,
and make a dichotomic measurement of $A_x$ an $C_z$ with the outcomes $a,c=\pm 1$. The observer $B$ has only one setting and the outcome $b=0,1,2,3$.

In the quantum mechanics based on real numbers, the separability between $P$ and $Q$ takes place in real space, which leads to tighter bounds on correlations
than in full complex space. The witness to distinguish the two cases reads
\begin{equation}
F=\sum_{x,z,b}(-1)^{s_{bx}}f_{xz}\langle A_xC_z||b\rangle,\label{fff}
\end{equation}
where the correlation is expressed in terms of probabilities as
\begin{equation}
\langle AC||b\rangle=\sum_{ac}acp(a,b,c)
\end{equation}
with coefficients $f_{jk}$ and
\begin{equation}
s_{bx}=\left\{\begin{array}{ll}
-1&\mbox{ for }b\neq x,\\
+1&\mbox{ otherwise. }
\end{array}\right.
\end{equation}

The goal of our research is to find the matrix $f$ such that the bound on $F$ is the lowest possible in the real case in comparison with the
maximum in the complex space.
In the classical case, the maximum reads
\be
F_c=\mathop{\mathrm{max}}_{|s_x|=|t_z|=1}\sum_{xz}f_{xz}s_xt_z
\ee

In the quantum case the maximum $F_q$ depends much on the actual form of the matrix $f$.
In \cite{cvr}, the authors constructed $f_{xz}$ for $x=1,2,3$, $z=1,2,3,4,5,6$ for the combination of three Clauser-Horne-Shimony-Holt Bell inequalities
\cite{bell,chsh,chsh31,chsh32}.

Here we consider the case of $4$ settings for the observer $C$, and a family of $f$s, where the complex quantum maximum, Tsirelson bound \cite{tsir}, can be found algebraically
and it is realized in the discussed setup. Namely let us take
\be
f=\begin{pmatrix}
\alpha&\beta&\gamma&q\\
\gamma&\alpha&\beta&q\\
\beta&\gamma&\alpha&q\end{pmatrix}
\ee
where
\be
q=-\sqrt{3}(\alpha\beta+\beta\gamma+\gamma\alpha)
\ee
with the constraints $\alpha^2+\beta^2+\gamma^2=1$ and $q\geq 0$.
We will show that 
\be
F_q=3+\sqrt{3}q.
\ee
Note that the following quantity is positive in both quantum and classical case,
\ba
&&q(\sqrt{3}C_4-A_1-A_2-A_3)^2/\sqrt{3}+\nonumber\\
&&(C_1-\alpha A_1-\beta A_2-\gamma A_3)^2+\nonumber\\
&&(C_2-\alpha A_2-\beta A_3-\gamma A_1)^2+\nonumber\\
&&(C_3-\alpha A_3-\beta A_1-\gamma A_2)^2.\label{sss}
\ea
Opening brackets and taking into account that $A^2=C^2=1$ we get
\be
\sum_{x,z}f_{xz}\langle A_xC_z\rangle\leq F_q
\ee
This maximum holds also if $B$ is included.

The example realizing the above maximum is constructed as follows.
We recall standard conventions in Appendix \ref{apa}. 
The working space consists of 4 qubits $APQC$ with $PQ\to B$.
The source states read
\be
\rho_L=(1-\boldsymbol \sigma^A\cdot\boldsymbol\sigma^P)/4,\:\rho_R=(1-\boldsymbol \sigma^C\cdot\boldsymbol\sigma^Q)/4
\ee
using shorthand notation $1\otimes 1\equiv 1^{\otimes 2}\equiv 1$ (a tensor product of identities).
The matrices $\sigma$ act on the qubit in the subscript while identity on the other qubits.
We assume projective measurements of $A$, $B$, $C$.
Four measurements at $B$ are defined by the projections:
\ba
&&B_0=(1-\boldsymbol \sigma^P\cdot\boldsymbol \sigma^Q)/4,\\
&&B_1=(1-\sigma^P_1\sigma^Q_1+\sigma^P_2\sigma^Q_2+\sigma^P_3\sigma^Q_3)/4,\nonumber\\
&&B_2=(1-\sigma^P_2\sigma^Q_2+\sigma^P_3\sigma^Q_3+\sigma^P_1\sigma^Q_1)/4,\nonumber\\
&&B_3=(1-\sigma^P_3\sigma^Q_3+\sigma^P_1\sigma^Q_1+\sigma^Q_2\sigma^Q_2)/4.\nonumber
\ea
After the the measurement at $B$, in the $AC$ basis:
\ba
&&\mathrm{Tr}_{PQ}B_0\rho_R\rho_R=(1-\boldsymbol \sigma^A\cdot\boldsymbol \sigma^C)/16,\\
&&\mathrm{Tr}_{PQ}B_1\rho_L\rho_R=(1-\sigma^A_1\sigma^C_1+\sigma^A_2\sigma^C_2+\sigma^A_3\sigma^C_3)/16,\nonumber\\
&&\mathrm{Tr}_{PQ}B_2\rho_L\rho_R=(1-\sigma^A_2\sigma^C_2+\sigma^A_3\sigma^C_3+\sigma^A_1\sigma^C_1)/16,\nonumber\\
&&\mathrm{Tr}_{PQ}B_3\rho_L\rho_R=(1-\sigma^A_3\sigma^C_3+\sigma^A_1\sigma^C_1+\sigma^A_2\sigma^C_2)/16.\nonumber
\ea

For observables $A=\boldsymbol a\cdot\boldsymbol \sigma^A$, $C=\boldsymbol c\cdot\boldsymbol \sigma^C$
with $\boldsymbol a\cdot\boldsymbol a=\boldsymbol c\cdot\boldsymbol c=1$
we have for the measurement outcome $a\to\pm 1$ (projections $(1\pm A)/2$), $b\to 0,1,2,3$ (projections $B_b$),  $c\to\pm 1$ (projections $(1\pm C)/2$), and correlations
\ba
&&\langle AC||b\rangle=\mathrm{Tr}ACB_b\rho_L\rho_R,\nonumber\\
&&
\langle AC||0\rangle=-\boldsymbol a\cdot\boldsymbol c/4,\\
&&
\langle AC||1\rangle=(a_2c_2+a_3c_3-a_1c_1)/4,\nonumber\\
&&\langle AC||2\rangle=(a_3c_3+a_1c_1-a_2c_2)/4,\nonumber\\
&&\langle AC||3\rangle=(a_1c_1+a_2c_2-a_3c_3)/4.\nonumber
\ea

We emphasize that the above formulas are valid in full complex space because of the complex matrix $\sigma_2$.

\section{The test of complex versus real quantum mechanics}

The maximum of $F$ in the complex quantum mechanics is obtained for $(a_x)_i=\delta_{ix}$, and 
\ba
&&\g{c}_1=-(\alpha,\beta,\gamma),\;\g{c}_2=-(\gamma,\alpha,\beta),\nonumber\\
&&\g{c}_3=-(\beta,\gamma,\alpha),\:\g{c}_4=(1,1,1)/\sqrt{3}\label{ccc}
\ea

\begin{figure}
\includegraphics[scale=.5]{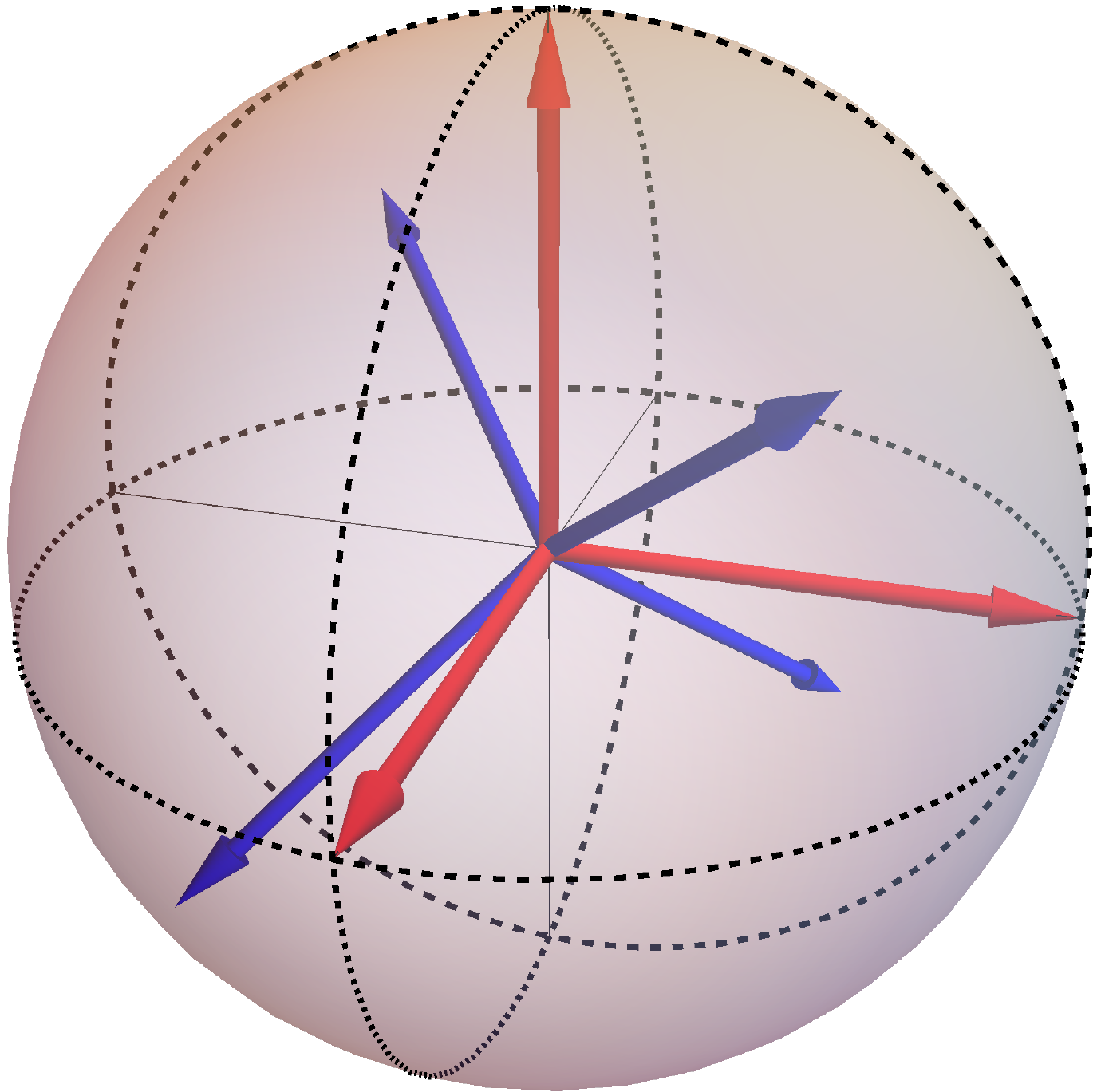}
\caption{The directions of the settings of $A$ (red, light) and $C$ (blue, dark) in the Bloch sphere, maximizing the $F_q/F_r$ ratio for $4$ settings of $C$
and $F$ defined by (\ref{fff})}\label{tet}
\end{figure}

In this case it is also possible to prove the impossibility of achieving this maximum in the real case. To see this note, that the maximal case,
saturating the inequality, in order to get zero from each of the squares in (\ref{sss}) must satisfy
\ba
&&\sqrt{3}C_4=A_1+A_2+A_3,\nonumber\\
&&C_1=\alpha A_1+\beta A_2+\gamma A_3,\nonumber\\
&&C_2=\alpha A_2+\beta A_3+\gamma A_1,\\
&&C_3=\alpha A_3+\beta A_1+\gamma A_2\nonumber
\ea
when acting on the state. Defining
\ba
&&C_1=\alpha C'_1+\beta C'_2+\gamma C'_3,\;C_2=\alpha C'_2+\beta C'_3+\gamma C'_1,\nonumber\\
&&
C_3=\alpha C'_3+\beta C'_1+\gamma C'_2,
\ea
we quickly find, using the fact $A_j^2=1=C_j^2$, that acting on the state,
\be
C'_j=A_j,\: A_iA_j=-A_jA_i,\; C'_iC'_j=-C_j'C_i'
\ee
for $i\neq j$. Using these identities one can repeat the reasoning in \cite{cvr} to show that the initial state $\rho_{APQC}$ is not real separable into
spaces $AP$ and $QC$. The actual maximum in the real case $F_r$ can be bounded from above by the numerical code (see later). It turns out that 
the maximal ratio $F_q/F_r>1$ for all choices  of $\alpha,\beta,\gamma$ but its maximum $\simeq 1.07$ is obtained for $\beta=\gamma=-\alpha=1/\sqrt{3}$, giving $F_q=4$ while $F_r=3.7367$ and $F_c=2\sqrt{3}$.
The directions $C$ lie in the vertices of the regular tetrahedron, see Fig.\ref{tet}, and our complex maximum coincides with Platonic (elegant) 
Bell inequalities \cite{chsh31,eleg1,platon}.

In the case of 3 settings for the observer $C$, the problem to find the maximum is as hard as the Tsirelson bound for $I_{3322}$ inequality \cite{I3322}.
Therefore we choose a family of states for a general form of $f$. The settings for $A$ are the same as in before.
For $C$ we take
\begin{equation}
 (c_z)_i=-\frac{f_{iz}}{\sqrt{f_{1z}^2+f_{2z}^2+f_{3z}^2}},
 \end{equation}
leading to the value
\be
F_q=\sum_z\sqrt{f_{1z}^2+f_{2z}^2+f_{3z}^2}.
\ee
To find the real maximum  $F_r$ in the case of four and three settings for $C$ we had to adopt the MATLAB script using the numerical technique of minimizing $F$ with semipositive matrices
of correlations of monomials of $A_x$ and $C_z$ up to given degree $n_A$ and $n_C$ \cite{mono}, based
on the fact that the real states are separable, i.e. $\rho_{LR}=\rho_L\rho_R$ if and only if $\rho_{LR}=(\rho^T_L)_R$, i.e. the state is equal its 
partial transform \cite{ppt}. Note that this is a stricter criterion than in the complex case, where separable states must have a positive partial transform
but not always  vice versa (not if and only if) \cite{pptc1,pptc2,pptc3}. Such a numerical problem is convex and a solution can be obtained using available 
semidefinite algorithms \cite{sdp} and tools \cite{mos,yal}. The script, being the modification of the original one from \cite{cvr}, is in Supplemental Material \cite{sup}.

\begin{figure}
\includegraphics[scale=.31]{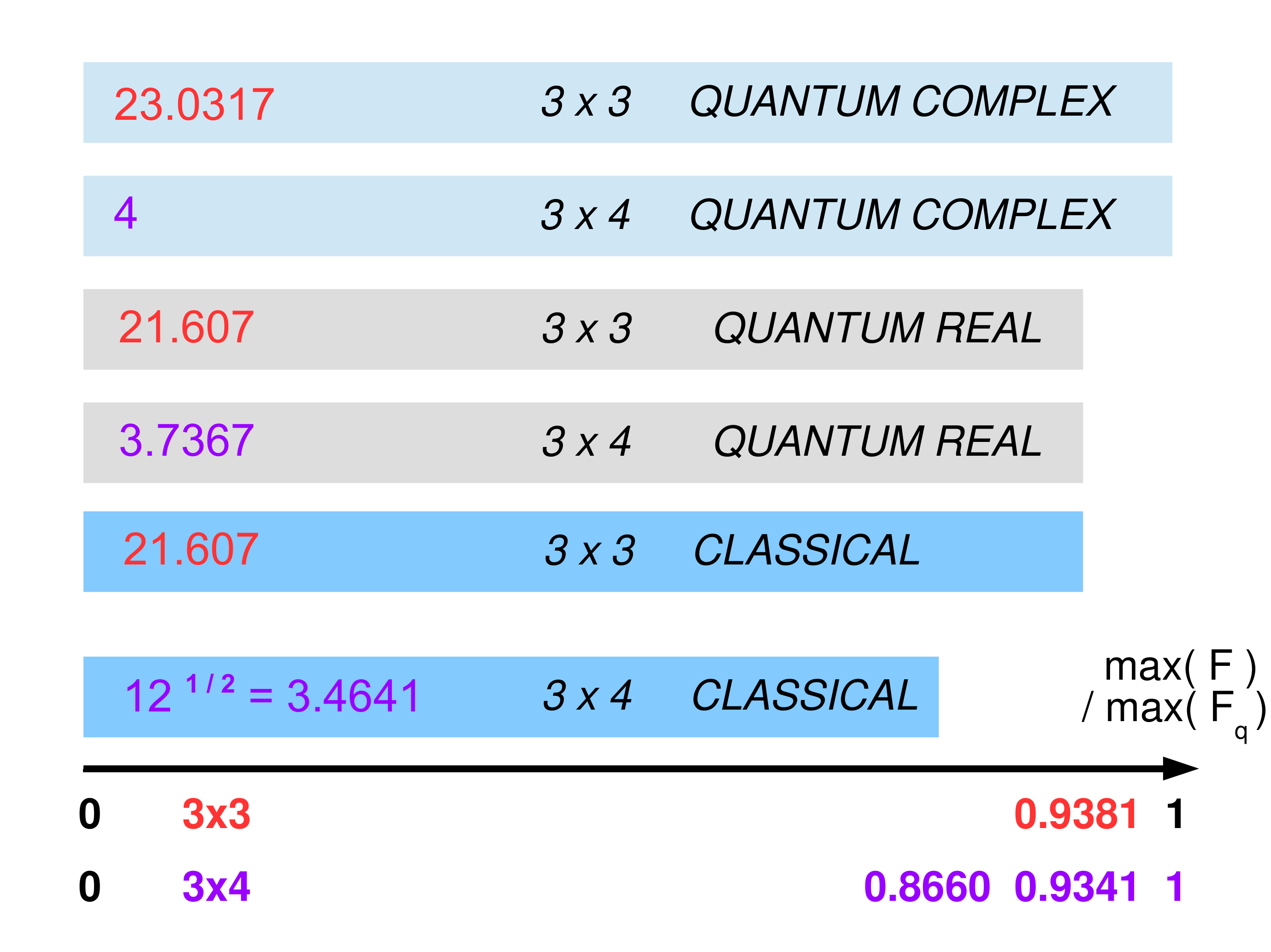}
\caption{The classical, quantum real and complex bound, $F_c$, $F_r$, $F_q$, respectively, and their ratio in the case of tests with either $3\times 4$ or $3\times 3$ 
settings for observers $A\times C$.}\label{sch}
\end{figure}

As an example, take $f$ equal
\be
\begin{pmatrix}
-2&3&3\\
3&-2&3\\
3&3&-2\end{pmatrix}.
\ee
Then $F_c=12$, $F_q=3\sqrt{22}\simeq 14.07$ while $F_r=13.677$ (taking $n_A=n_C=2$) so there is clearly a quantum state violating the real bound.
Interestingly, we could also run $n_A=n_C=3$ case and the value does not change (up to machine precision).
Unfortunately, the ratio $F_q/F_r\simeq 1.029$ is quite demanding experimentally so we run numerical scan through a wide range of matrices $f$, additionally
boosting the largest value by the steepest descent method.
We found the maximal ratio $F_q/F_r\simeq 1.06594$ for $f$ equal
\be
\begin{pmatrix}
3& -5.009& -4.99\\
-5.01& 2.6& -5.09\\
-5.11& -5& 3\end{pmatrix}
\ee
with $F_r=F_c=21.607$, $F_q=23.031685$. The coincidence of the real and classical maximum indicates that the constraints of the numerical method
produce here the exact bound already for $n_A=n_C=2$. We illustrated the bounds in the case of 3 and 4 settings for $C$ in Fig. \ref{sch}.

Reduction to two settings and two outcomes for $A$ or $C$ gives equal real and complex bounds $F_q=F_r$ for maximal entanglement. 
For both $A$ and $C$ with two settings and outcomes, 
we confirmed the general equality $F_q=F_r$ by a numerical survey on a sample of 40000 random test points, see Appendix \ref{apb}.
Relaxing more assumptions, e.g. one observer has more settings or we allow more outcomes, seems highly nontrivial and we cannot make any definite conclusions yet.

\section{Conclusions}

We have found tests of real separability in contrast to complex separability alternative to the recently found in Ref. \cite{cvr}. Our proposals
require fewer settings for the observers but the ratio between complex and real bound is also lower. The numerical check takes a much shorter time.
There are still some open questions: despite extensive numerical search, we could not reduce the test to two setting for observers $A$ or $C$, nor could provide a 
mathematical proof of impossibility. Also, when the real maximum is higher than the classical one, a numerical search through
low-dimensional real quantum systems could not find any example above the classical bound. It may probably require many more dimensions, just like the $I_{3322}$ case \cite{I3322}.
One can explore possible inequalities based on more settings (six for $A$ and $C$) using other Platonic inequalities \cite{platon}, specially to
increase $F_q/F_r$ ratio, but 
it  can be demanding numerically. Another interesting problem is to check three or more outcomes and all cases where one party has only two settings,
whether it is possible to rule these cases out by arguments similar to Bell tests \cite{fine}.
 Finally, we emphasize that such tests do have loopholes -- the sources $P$ and $Q$ can be already entangled, 
and the observers could communicate during measurements unless one uses a spacelike regime as in recent Bell tests \cite{hensen,nist,vien,munch}.
The communication can be checked in every such experiment, including the already existing ones \cite{cvre1,cvre2}, verifying no-signaling i.e.
independence of
$\langle A_x||b\rangle$, $\langle C_z||b\rangle$, and $\langle 1||b\rangle\equiv p(b)$, of $z$, $x$ and $xz$, respectively, 
but the published data are insufficient for these specific claims.

\section*{Acknowledgements}

J. B. acknowledges fruitful discussions with J. Rossell\'o, 
M. del Mar Batle and R. Batle. A.B. acknowledges discussion with J. Tworzyd{\l}o.

\appendix
\section{Notation}
\label{apa}

We use Pauli matrices
\be
\sigma_1=\begin{pmatrix}
0&1\\
1&0\end{pmatrix},\:\sigma_2=\begin{pmatrix}
0&-i\\
i&0\end{pmatrix},\:\sigma_3=\begin{pmatrix}
1&0\\
0&-1\end{pmatrix},\:
1=\begin{pmatrix}
1&0\\
0&1\end{pmatrix},
\ee
with $\sigma_a\sigma_b=\delta_{ab}+i\epsilon_{abc}\sigma_c$, summation convention, i.e. $X_aY_a\equiv \sum_{a=1,2,3}X_aY_a=\boldsymbol X\cdot\boldsymbol Y$,
and 
\ba
&&\delta_{ab}=\left\{\begin{array}{ll}
1&\mbox{ for }a=b\\
0&\mbox{ for }a\neq v
\end{array}\right.,\nonumber\\
&&
\epsilon_{abc}=\left\{\begin{array}{ll}
+1&\mbox{ for }abc=123,231,312,\\
-1&\mbox{ for }abc=321,213,132,\\
0&\mbox{ otherwise }
\end{array}\right..
\ea
We also use  identities  
\ba
&&\epsilon_{abc}\epsilon_{ade}=\delta_{bd}\delta_{ce}-\delta_{be}\delta_{cd},\nonumber\\
&&\epsilon_{abc}\epsilon_{abd}=2\delta_{cd},\:\delta_{aa}=3,\:\delta_{ab}\delta_{bc}=\delta_{ac}.
\ea

\section{No-go with two settings and two outcomes}
\label{apb}

Here we prove that every test involving two settings and two outcomes for both observers will give $F_q=F_r$.
First, we show that $A_{1,2}$ are real matrices in some basis.
The states $AP$ and $QC$ can be written in terms of diagonal Schmidt decomposition.
By convexity we can also project $A_j$ onto the space spun by nonzero Schmidt elements and assume $A_j^2=1$.
We can start from a diagonal basis of $A_1$, where $A_1=\mathrm{diag}(1,..,1,-1,..,-1)$.
We also adjust this basis so that
\begin{equation}
A_2=\begin{pmatrix}
A_+&A'\\
A^{\prime\dag}&A_-
\end{pmatrix},
\end{equation}
where $A_{\pm}$ are diagonal in the respective subspaces of $A_1=\pm 1$. Since $A_2^2=1$, we get
\begin{equation}
A_+A'+A'A_-=0.
\end{equation}
It means that either $(A_0)_{jk}=0$ or $A_{+j}+A_{-k}=0$. In the latter case, we can group subspaces $A_{+j}=-A_{-k}=a$.
All off-diagonal elements of $A_0$ between different values $a$ must be zero. Within the particular subspace $a$ we can make singular value decomposition of $A'|_a$ resulting in its diagonal
form (appended by zeros for non-square $A'$). By phase tuning, we end up in a real matrix. Note that the construction leads to splitting of the whole space into
trivial 1-dimensional spaces where $A_1=\pm 1$ and $A_2=\pm 1$ or 2-dimensional ones, where $A_1$ and $A_2$ are some combinations of $\sigma_1$ and $\sigma_3$.
These spaces are not connected by elements of $A_j$.

If both $A_j$ and the Schmidt decomposition 
\begin{equation}
|L\rangle=\sum_j \lambda_j|jj\rangle,\:\sum_j|\lambda_j|^2=1,
\end{equation}
are real valued in the same basis as $A_j$ are, we can use the following reasoning.
All monomials consisting of $A_j$ have the form
\begin{equation}
m(A_1,A_2)=\cdots A_1A_2A_1A_2\cdots
\end{equation}
i.e. $A_1$ at odd positions and $A_2$ at even ones, or viceversa. In any case, the mononomial is a real matrix in the constructed basis. Therefore
\begin{eqnarray}
&&\langle m(A_1,A_2)\rangle=\langle m^\dag(A_1,A_2)\rangle^\ast=\nonumber\\
&&\langle m^T(A_1,A_2)\rangle^\ast=\langle m^T(A_1,A_2)\rangle
\end{eqnarray}
where the transpose is in the real basis. It is true whenever the state has also real representation.
Now the monomial matrix, 
\begin{eqnarray}
&&G_b(m(A_1,A_2),m(C);m'(A_1,A_2),m'(C))=\nonumber\\
&&\langle m^T(A_1,A_2)m'(A_1,A_2)m^\dag(C)m'(C)B_b\rangle,
\end{eqnarray}
is complex semidefinite in general but satisfies
\begin{eqnarray}
&&G=\sum_b G_b(m(A_1,A_2),m(C);m'(A_1,A_2),m'(C))=\nonumber\\
&&\langle m^T(A_1,A_2)m'(A_1,A_2)m^\dag(C)m'(C)\rangle=\\
&&\langle m^{\prime T}(A_1,A_2)m(A_1,A_2)m^\dag(C)m'(C)\rangle,\nonumber
\end{eqnarray}
i.e. equality with the partial transpose. Taking $G_b\to (G_b+G_b^\ast)/2$ one obtains a real semipositive matrix.
Therefore the algorithm from \cite{cvr} will give the same bound for the real and complex states.
For the maximally entangled states,
$\lambda_j$ are independent of $j$. The basis transformations $U$ on $A$ spaces can be then moved through the constant onto $P$.

If $C$ has also only two settings and outcomes we reduce it to a real representation analogously. 
For nonmaximally entangled states, taking into account splitting $A_j$ and $C_j$ into 2-level spaces,
by convexity we can reduce the problem to a 2-level space for $A$ and $C$ (and $P$ and $Q$ by Schmidt decomposition).
However, for a nonmaximally entangled states we cannot push $U$ through the states as they are not necessarily diagonal.
Note that such states can reveal nonclassical features in special situations, where it is impossible with the maximally entangled ones \cite{abadp}.
Since the $PQ$ space has $2\times 2=4$ dimensions, we can have maximally $4$ outcomes for $B$. 
In order to deal with this situation, we have resorted to a numerical exploration. We have generated 40000 random sets of Bell-type 36 parameters 
$f_{bxz}$ for quantity
\begin{equation}
F=\sum_{bxz}f_{bxz}\langle A_xC_z||b\rangle,\label{maxac}
\end{equation}
with $i=0123$, $x,z=012$ and $A_0=1=C_0$, in the hypercube $[-1,1]^{36}$, and, for each one of them, computed the maximum value.

\begin{figure*}
\includegraphics[scale=0.3]{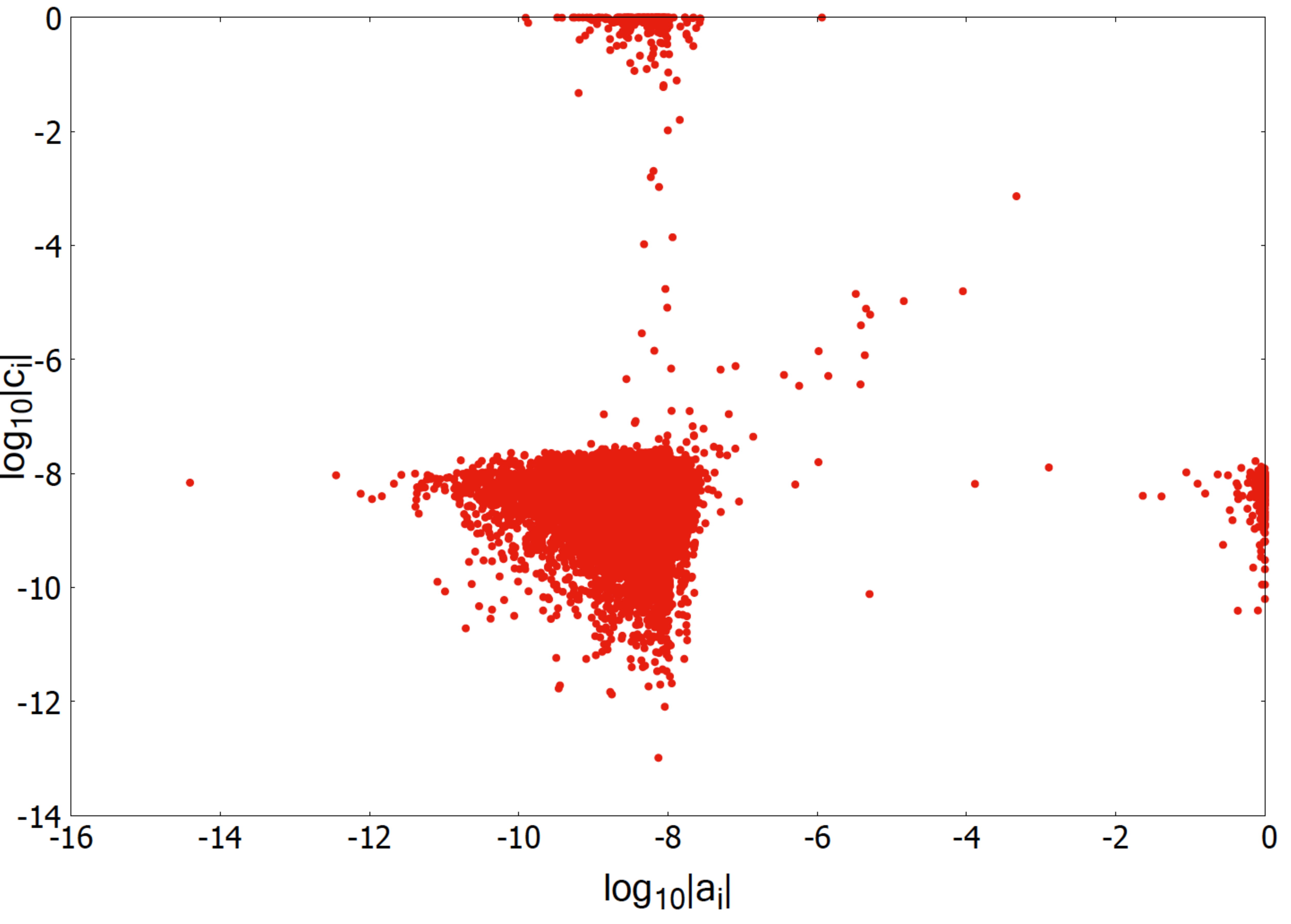}
\caption{Points $(a_i,c_i)$ in the numerical test of 40000 random points maximizing (\ref{maxac}), see text. Only simultaneous $a_i,c_i\sim 1$ would give a complex
example, irreducible to real space.}\label{cva}
\end{figure*}

The task is feasible, but going further to higher number of sets is computationally demanding.   
The states and observables can be then represented as follows.
Diagonal and real $U$, $V$, $B_b=|b\rangle\langle b|$ where 
\begin{equation}
|b\rangle=\sum h_{bjk}|jk\rangle,
\end{equation}
for the bases $|0\rangle$,$|1\rangle$ of $P$ and $Q$
being rows of the unitary $4\times 4$ matrix $H$, i.e. $H_{jk}=h_{j,2j+k}$, which can be generated by the product of 6 matrices
\begin{equation}
H^{(JK)}_{jk}=\left\{\begin{array}{ll}
\cos\theta_{JK}&\mbox{ for }j=k=J,K\\
e^{i\psi_{JK}}\sin\theta_{JK}&\mbox{ for }j=J,k=K\\
-e^{-i\psi_{JK}}\sin\theta_{JK}&\mbox{ for }j=K,k=J\\
1&\mbox{ for }j=k\neq J,K\\
0&\mbox{ otherwise}
\end{array}\right.
\end{equation}
with $J<K$. We check if the maximal solution  retrieves nonzero values for $a_{12}=-a_{22}=a_i$ and $c_{12}=-c_{22}=c_i$ (the matrices can be always put into this form
by a phase shift in qubit space).
The final optimal values for $a_i$, $c_i$ are depicted in Fig. \ref{cva}. There are two types of data, namely, 
the ones with either $|a_i|$ or $|c_i|$ close to unity, and the rest being very small. 
During our numerical survey, we have not encountered any case with both $|a_i|$ and $|c_i|$ any near to one 
(our criterion has been $|a_i|>0.01$ and $|c_i|>0.01$). We cannot rule out entirely 
the possibility of finding counterexamples that refute our main claim, but our results indicate that it is unlikely.

\end{document}